\documentclass{icrc2009}

\usepackage{graphicx}   
\usepackage[caption=false]{caption}    
\usepackage[font=footnotesize]{subfig} 
\usepackage{fixltx2e}
\usepackage{url}

\usepackage[usenames,dvipsnames]{color}
\usepackage{hyperref}
\hypersetup{
   colorlinks,%
   citecolor=RoyalBlue,%
   filecolor=RedViolet,%
   linkcolor=ProcessBlue,%
   urlcolor=RedViolet
}

\newcommand{\shorttitle}[1]%
{\markboth{Proceedings of the 31\MakeLowercase{$^{st}$} ICRC, {\L}\'{o}d\'{z} 2009}{#1} }
\newcommand{\etal}{\MakeLowercase{\textit{et al. }}} 

\begin{document}
\title{Water Cherenkov Detectors response to a Gamma Ray Burst in the Large
Aperture GRB Observatory}

\author{ \vbox{ D.~Allard$^a$,
C.~Alvarez$^b$,
H.~Asorey$^c$,
H.~Barros$^d$,
X.~Bertou$^c$,
M.~Castillo$^e$,
J.M.~Chirinos$^f$,
\underline{A.~De Castro}$^d$,
S.~Flores$^g$,
J.~Gonzalez$^h$,
M.~Gomez Berisso$^c$,
J.~Grajales$^e$,
C.~Guada$^i$,
W.R.~Guevara Day$^j$,
J.~Ishitsuka$^g$,
J.A.~L\'opez$^k$,
O.~Mart\'inez$^e$,
A.~Melfo$^i$,
E.~Meza$^l$,
P.~Miranda Loza$^m$,
E.~Moreno Barbosa$^e$,
C.~Murrugarra$^d$,
L.A.~N\'u\~nez$^i$,
L.J.~Otiniano Ormachea$^j$,
G.~Perez$^n$,
Y.~Perez$^i$,
E.~Ponce$^e$,
J.~Quispe$^m$,
C.~Quintero$^i$,
H.~Rivera$^m$,
M.~Rosales$^i$,
A.C.~Rovero$^o$,
O.~Saavedra$^p$,
{H.~Salazar}$^{e,1}$,
J.C.~Tello$^d$,
R.~Ticona Peralda$^m$,
E.~Varela$^e$,
A.~Velarde$^m$,
L.~Villasenor$^q$,
D.~Wahl$^g$,
M.A.~Zamalloa$^r$
 (LAGO Collaboration)}\\
{$^a$}APC, CNRS et Universit\'e Paris 7. France\\
{$^b$}Universidad Autonoma de Chiapas, UNACH.  M\'exico \\
{$^c$}Centro At\'omico Bariloche, Instituto Balseiro.  Argentina \\
{$^d$}Laboratorio de F\'isica Nuclear, Universidad Sim\'on Bol\'ivar,
Caracas. Venezu
ela \\
{$^e$}Facultad de Ciencias F\'isico-Matem\'aticas de la BUAP.  M\'exico \\
{$^f$}Michigan Technological University. USA\\
{$^g$}Instituto Geofisico del Per\'u, IGP. Lima - Per\'u\\
{$^h$}Universidad de Granada. Spain\\
{$^i$}Universidad de Los Andes, ULA. M\'erida - Venezuela\\
{$^j$}Comisi\'on Nacional de Investigaci\'on y Desarrollo
Aeroespacial, CONIDA. San I
sidro Lima - Per\'u\\
{$^k$}Universidad Central de Venezuela, Facultad de Ciencias,
Departamento de F\'isic
a. Venezuela\\
{$^l$}Universidad Nacional de Ingenieria, UNI. Lima 25 - Per\'u\\
{$^m$}Instituto de Investigaciones F\'isicas, UMSA. Bolivia \\
{$^n$}Universidad Polit\'ecnica de Pachuca.  M\'exico \\
{$^o$}Instituto de Astronom\'ia y F\'isica del Espacio. Argentina \\
{$^p$}Dipartimento di Fisica Generale and INFN, Torino. Italy\\
{$^q$}Instituto de F\'isica y Matem\'aticas, Universidad de
Michoac\'an. M\'exico\\
{$^r$}Universidad Nacional San Antonio Abad del Cusco. Per\'u
}


\shorttitle{A. De Castro \etal WCD response to GRB at the LAGO}
\maketitle

\begin{abstract}
In order to characterise the behaviour of Water Cherenkov Detectors (WCD)
under a sudden increase of 1 GeV - 1 TeV background photons from a Gamma
Ray Burst (GRB), simulations were conducted and compared to data acquired
by the WCD of the Large Aperture GRB Observatory (LAGO).  The LAGO operates
arrays of WCD at high altitude to detect  GRBs using the single particle technique. The LAGO sensitivity to GRBs is derived from the reported simulations of the gamma initiated particle showers in the atmosphere
and the WCD response to secondaries.

\footnotetext[1]{presenting and corresponding author,
\href{mailto:adcastro@usb.ve}{adcastro@usb.ve}}

\end{abstract}

 
\section{Introduction}
Gamma Ray Bursts (GRBs) have captured the attention of the astrophysics community during the second half of the past century, and are nowadays still a hot topic of study. Instruments in satellites and ground based experiments around the planet have been designed in order to study this impressive phenomena (see \cite{Meszaros:2006rc, Kaneko:2006qe} and references therein). Large data set of GRB properties can be found in \cite{Kaneko:2006qe} for example.\\     
GRB emissions in the keV to MeV energy range has been studied in great detail. Though high energy GRB emissions have been reported (for example up to 2 GeV for GRB940217 seen by EGRET), there is little understanding on the subject. Satellite Fermi will be in charge of this mission and ground based experiments will also contribute. Theoretical predictions on the high energy emission in  relativistic fireball models are given in \cite{Meszaros:1993ft, Vietri:1997st, Baring:1997tv}. Feasibility of ground based observation of Gamma Ray Bursts in the $1$GeV and $1$ TeV energy range using the singe particle technique  has been extensively analysed (see for example  \cite{Castellina:2000jq, Vernetto:1999jz, Allard:2008zz, Bertou:2007hh}). The Large aperture Gamma-ray-bursts Observatory (LAGO) is designed to observe high energy GRB emissions by the single particle technique using water Cherenkov detectors (WCD) (see \cite{1413} in this volume). It consists of various sites, located around the world, with large GRB detection efficiency, either for their size or their altitude. The LAGO has 2 sites in operation, they are: Sierra Negra (4557 m a.s.l.), M\'exico and Chacaltaya (5200 m a.s.l.), Bolivia. One site at Pico Espejo (4765 m a.s.l.), Venezuela is at present under construction and new sites are now under consideration in Argentina, Per\'u, Colombia, Guatemala and Himalaya. High energy photons from a GRB are expected to produce particle cascades after reaching the atmosphere, thought with few secondary particles reaching the Earth surface. Nevertheless, many photons detectable by WCD are expected to arrive during a GRB.\\
In this paper we study the detection prospect of a GRB at three high altitude LAGO sites: Sierra Negra, Chacaltaya and Pico Espejo. In order to achieve this end, we simulated billions of showers initiated by photons with energies between $0.5$ GeV and $1$ TeV generated above each site, and the response of WCDs to secondaries reaching the ground level. 
 \begin{figure*}[!th]
   \centerline{\subfloat[]{
   \includegraphics[width=2.5in]{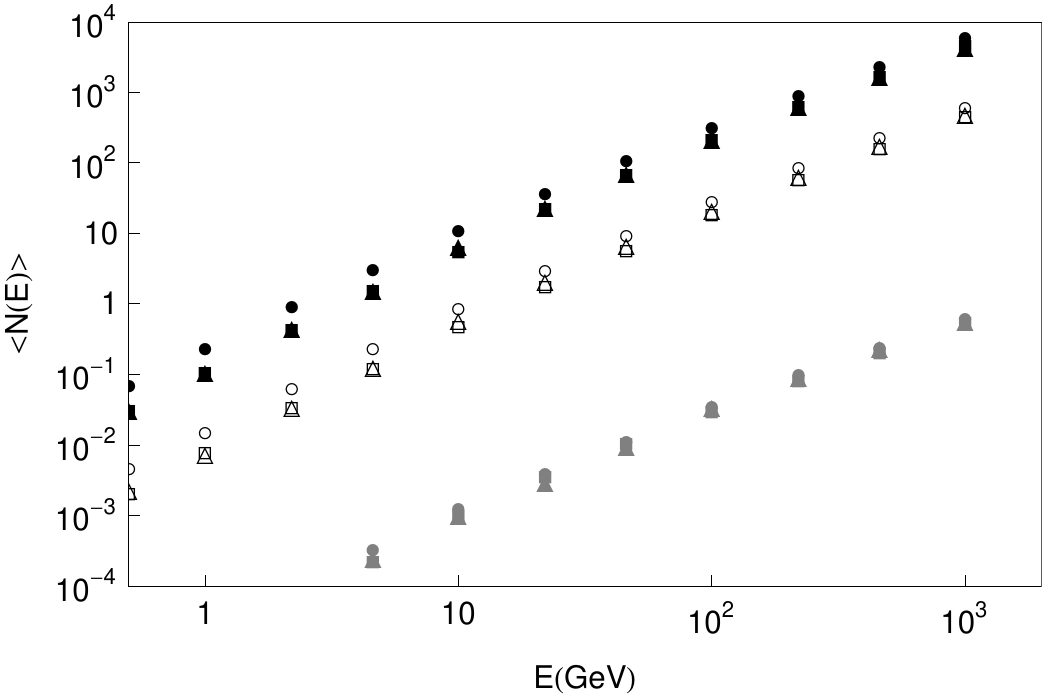} \label{fig:MNGepmuvsEp}}
              \hfil
              \subfloat[]{
              \includegraphics[width=2.5in]{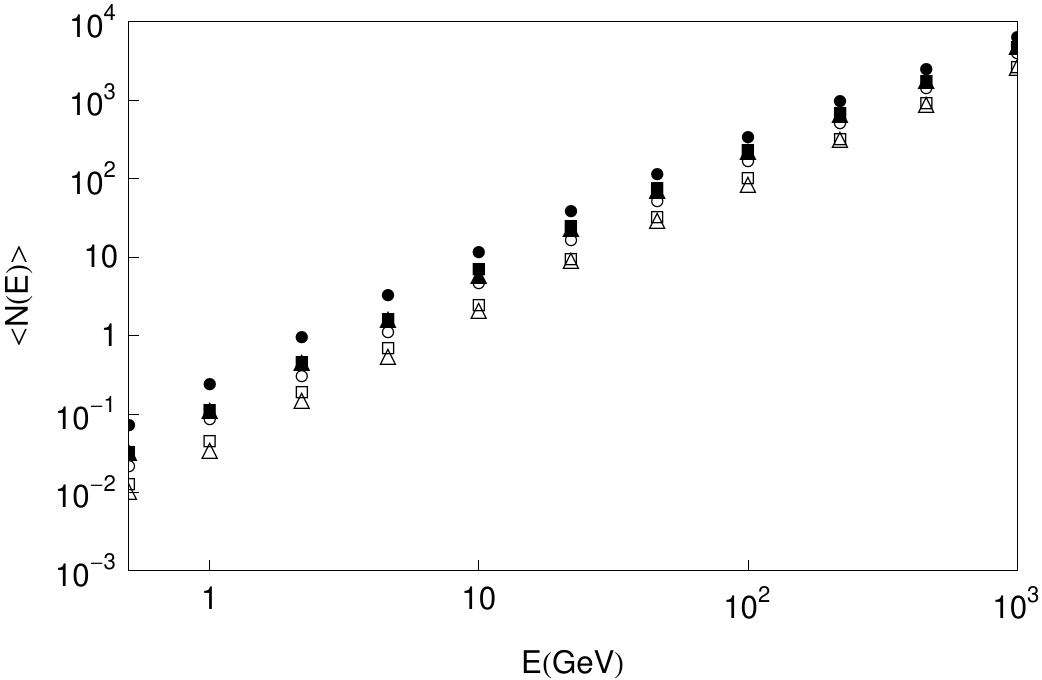} \label{fig:MNallvsEpZ}}
             }
   \caption{Left: Mean number of secondary photons (black), $e^+/e^-$ (white) and $\mu^+/\mu^-$ (gray), reaching the groung level vs primary energy. Right: Mean number of all secondary particles reaching the ground level vs primary energy for zenith angles $0^\circ$(black) and $30^\circ$(white). In both cases circles, triangles and squares stand for Chacaltaya, Pico Espejo and Sierra Negra respectively.}
   \label{double_fig}
 \end{figure*}

\section{Photon Showers at Ground Level\label{sec:Showers}}
In this section we investigate photon shower at ground level for three sites in the LAGO Observatory, Chacaltaya ($16^\circ 20^\prime$ S, $68^\circ 10^\prime$ W , $5200$ m a.s.l.), Bolivia; Sierra Negra ($18^\circ 59^\prime$ N, $97^\circ 18^\prime$  W , $4557$ m a.s.l.), M\'exico and Pico Espejo ($8^\circ 38^\prime$ N, $71^\circ 09^\prime$ W , $4765$ m a.s.l.), Venezuela. Primary energies are considered from $0.5$ GeV to $1$ TeV with zenith angles between $0^\circ$ to $30^\circ$. Showers were simulated using the Extensive Air Showers package AIRES\cite{aires}. No thinnig was applied, thus all particles were followed until absorbed in the atmosphere or reached the ground. We simulated showers with and without geomagnetic field and found no significant difference in the shower development at these latitudes. We analyse the nature, number and energy spectrum of secondary particles reaching the ground level as function of the photon primary energy and zenith angle of injection.\\ 
\begin{table}[!h]
  \caption{}
  \label{table:indice}
  \centering
  \begin{tabular}{|c|c|c|c|}
  \hline
     &  Chacaltaya & Pico Espejo&Sierra Negra \\
   \hline 
$\gamma$ & 1.5 &1.6 & 1.6\\
    $e^+/e^-$ & 1.6 & 1.6&1.6 \\
$\mu^+/\mu^-$ & 1.4 & 1.6&1.6 \\
   \hline
  \end{tabular}
  \end{table}
  \begin{table*}[th]
  \caption{}
  \label{table:prom100300}
  \centering
  \begin{tabular}{|c|c|c|c|}
  \hline
   &$\gamma$ (100 GeV/300 GeV) &$e^+/e^-$ (100 GeV/300 GeV)& $\mu^+/\mu^-$ (100 GeV/300 GeV)\\
   \hline 
   Chacaltaya   & 291/1259 &26/100 & 0.02/0.2\\
   Pico Espejo  & 197/1000 & 19/97 & 0.03/0.1\\
   Sierra Negra & 194/920 & 17/85 & 0.03/0.1 \\
   \hline
  \end{tabular}
  \end{table*}\\
In figure \ref{fig:MNGepmuvsEp} we represent the mean number of particles  (photons, $e^+/e^-$ and $\mu^+/\mu^-$) reaching the ground level as function of the primary photon energy at the three sites. The number of particles at ground level increases with the primary energy as a power law for the sites with index shown in table \ref{table:indice}. 
The mean number of secondary photons are about one order of magnitude above the mean number of $e^+/e^-$, while the number of muons start to be important after $\sim 1$ TeV for the three sites. Already some particles arrive for the lowest primary energies while hundreds and thousands of secondary photons arrive around $100$ GeV and $300$ GeV respectively (see table \ref{table:prom100300}). This is an advantage for surface detectors able to catch secondary photons.\\ We compare the development of vertical showers with the ones slanted at $30^\circ$ of zenith angles. The mean number of all secondary particles reaching the ground level vs primary energy for these two angles of injection at the three sites are displayed in figure\ref{fig:MNallvsEpZ}. Significant differences are observed for the three sites, which means that high altitude surface detectors are very sensible to the primary particle injection angle with respect to the detectors in the atmosphere. Next section is dedicated to the response of the surface detectors to vertical showers, studies depending on the zenith angle will be reported else where\cite{EnPreparacion} 
\begin{figure*}[!th]
  \centerline{\subfloat[]{
  \includegraphics[width=2.5in]{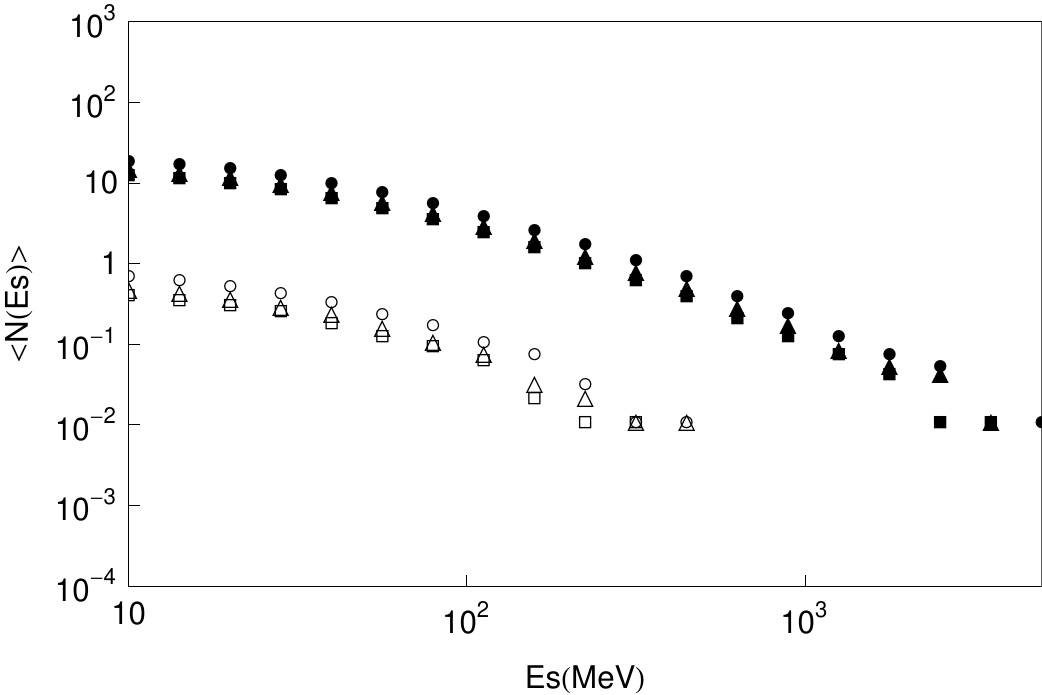}\label{fig:Spectrum100}}
  \hfil
  \subfloat[]{
  \includegraphics[width=2.5in]{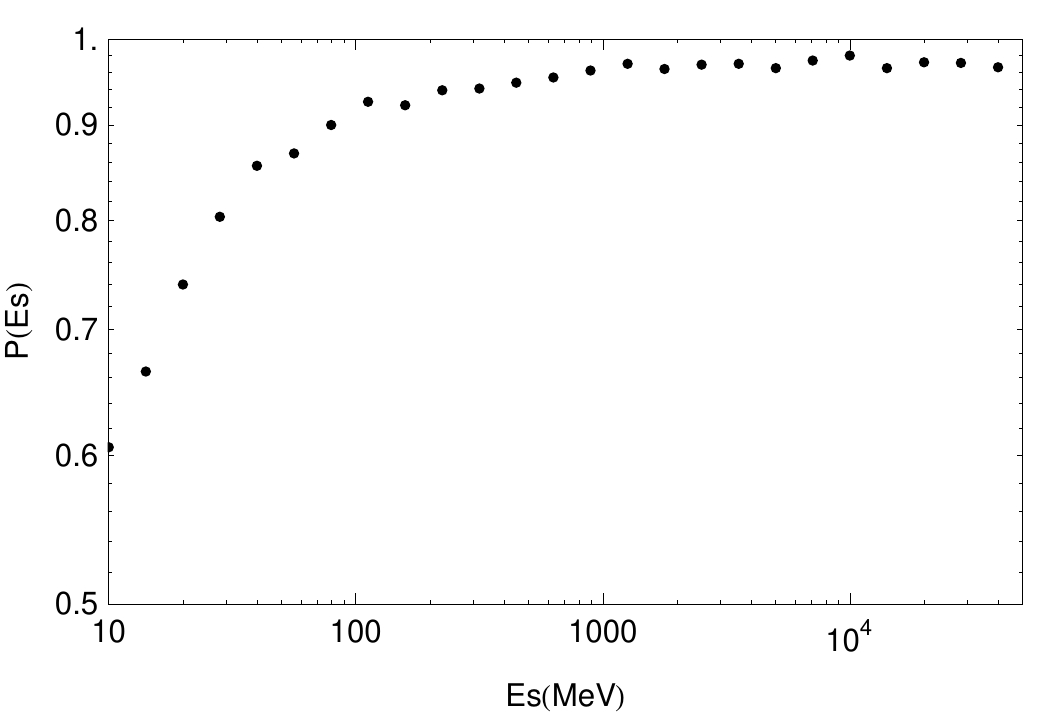}\label{fig:ProbDisparo}}}
  \caption{Left: Energy spectrum of secondary particles reaching the ground level al Chacaltaya(circles), Pico Espejo (triangles) and Sierra Negra (squares) for primary energies $10$ GeV (white) and $100$ GeV (black). Right: Probability of particle detection at a $4m^2$ WCD for energy range from $10$ MeV to $40$ GeV}
 \end{figure*}
\section{Water Cherenkov detectors response to high energy photon showers}
\begin{figure*}[!th]
   \centerline{\subfloat[]{
   \includegraphics[width=2.5in]{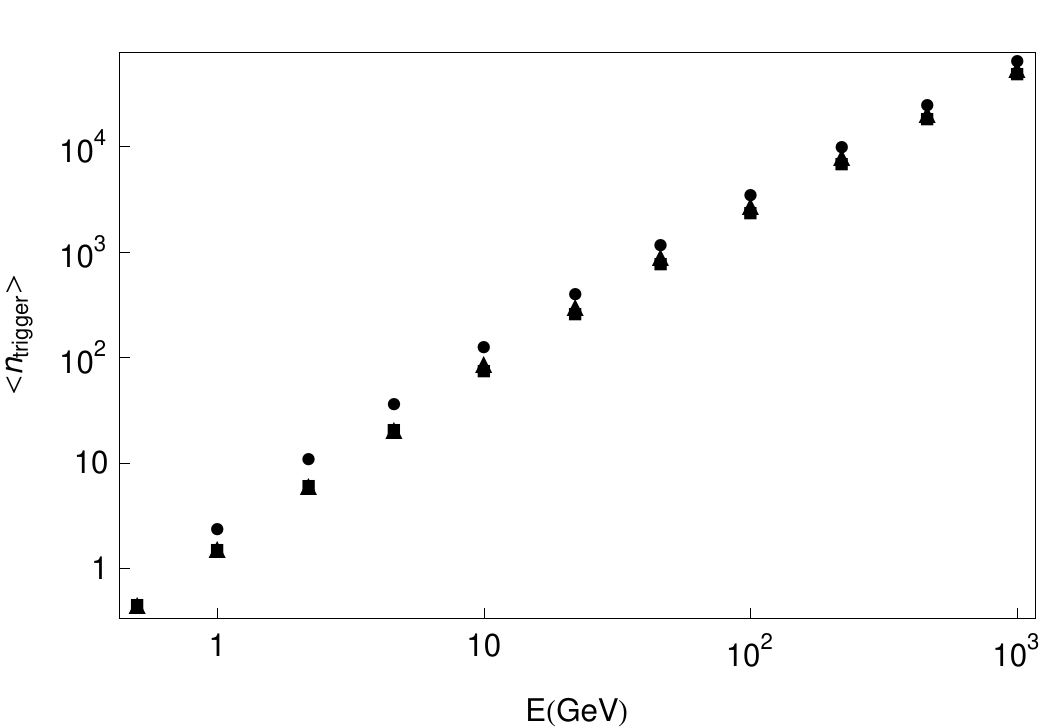} \label{fig:NtriggvsEp}}
              \hfil
              \subfloat[]{
              \includegraphics[width=2.5in]{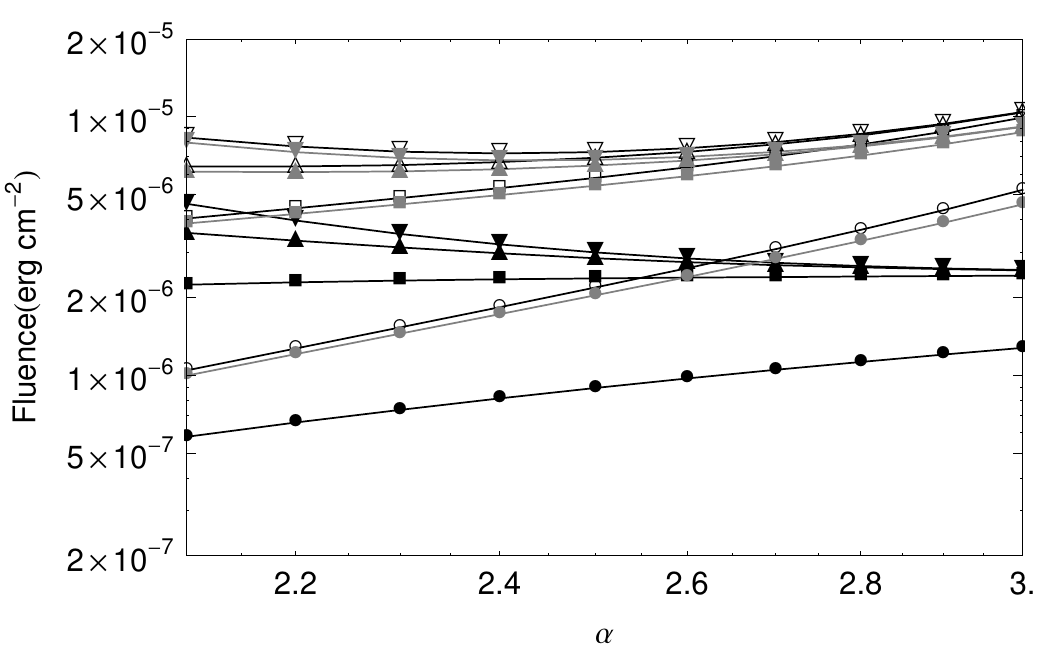} \label{fig:Fluencevsalpha}
              }
             }
   \caption{Left: Mean number of triggers as function of the primary energy for Chacaltaya (circles), Pico Espejo (triangles) and Sierra Negra (squares). Right: Minimal fluences vs the spectral parameter $\alpha$ for different $E_{max}$ at each site, black symbols stand for Chacaltaya, white for Pico Espejo and Gray for Sierra Negra. $E_{max}=1$GeV circles, $E_{max}=10$ GeV squares, $E_{max}=100$ GeV up triangles and $E_{max}=1$ TeV down triangles}
\end{figure*}

The LAGO aims to detect secondary particles from showers generated by photons using the single particle technique \cite{Vernetto:1999jz} which essentially involves the detection of secondary particle by individual WCDs. As photon showers around GeV give only isolated particles at ground level, even at high altitudes, it is not easy to reconstruct the shower development. Nevertheless, if secondary particles generated by these gamma rays represent a significant excess of events on the background a GRB would be detected in coincidence with satellites. The advantage of WCD relies in the fact that during a GRB  many photons are expect to arrive, most of them able to create pairs in the water tanks with sufficient energy for Cherenkov emission.\\
From our previous study on the features of the particles at ground level at each site, we are now able to analyze the response of a WCD to these secondary particles. Each site consists of 3  $\times (\sim 4$ m$^2)$ WCDs placed in an irregular array (there are no specifications in the separation of the tanks since we are using the single particle technique). For the case of these three LAGO sites, the efficiency to GRB detection is due to the high altitude of the sites \cite{Allard:2008zz} (details about the experimental set up can be found in \cite{1417}).\\            
We simulated the WCD response to secondary particles reaching the ground level, produced in showers initiated by high energy photons, for each high altitude LAGO site. In order to achieve this end, we used a fast WCDs simulator developed by the LAGO collaboration. The main parameters considered for the detectors are the tanks geometry, the water altitude; and the number, geometry and position of PMTs. We simulated photons, $e^+/e^-$ and $\mu^+/\mu^-$, entering vertically in the tanks\footnote{It has been reported in \cite{Bertou:2007hh} that the response of WCD does not depend much on the incidence angle of the secondary particles except at very low energies such that electrons can be absorbed by the plastic layer of the detector.      
}, for energies between $10$ MeV and $10$ GeV \footnote{AIRES output present histograms from $10$ MeV to $7079$ GeV}. Energy spectrum of secondary particles reaching the ground level for the three sites, produced in electromagnetic showers with $10$ GeV and $100$ GeV of primary energy is presented in figure \ref{fig:Spectrum100} as an example. Similar behavior is found for the three sites, while Chacaltaya highlights over the others two. For both primary energies we observe that most of the particles arrive with energies about $10$ MeV, then for $E=10$ GeV the spectrum presents a steepening after $\sim 14$ MeV and almost fade away over $0.2$ GeV whereas for $E=100$ GeV the steepening appears after $\sim 20$ MeV and we can still see particles over the $1$ GeV.      
The threshold energy for $e^+/e^-$ Cherenkov production in pure water is $\sim 0.77$ MeV making photons with $E_{\gamma} > 1.55$ MeV suitable to crate pairs detectable for our PMTs, thus our WCD response simulations are good enough in the energy spectrum of secondaries obtained with shower simulations.\\  
Detection probability $P(E_s)$ for each $4m^2$ tank is plotted in figure \ref{fig:ProbDisparo} which is defined as 
\begin{equation}
P(E_s)=\frac{n_{trig}}{N_{total}(E_s)}
\end{equation}
where $n_{trig}(E_s)$ is the total number of triggers and $N_{total}(E_s)$ is the total number of particles entering in the tank. We used 4 ADC as threshold for our trigger criteria. In this plot we can appreciate that detection probability rapidly increase between $E_s=10$ MeV and $E_s= 100$ and saturates near $\sim 1000$ MeV.\\ 
We calculated the mean number of triggers as function of the primary energy from the convolution of detection probability of particles hitting the WCD, and the differential spectrum of secondary particles, for each primary energy, taken from the shower simulations. 
Results for vertical showers are shown in figure \ref{fig:NtriggvsEp}. In this plot we can observe an increment of the mean number of triggers with the primary energy, being Chacaltaya clearly more sensitive to low energy photons than Pico Espejo and Sierra Negra (both overlapped along their curves). 

\section{GRB detection at high altitude LAGO sites} 
The LAGO is designed to detect GRBs by a sudden increase of the background in a short period of time. That means the counting rate in a surface observatory, summed over all the tanks, should be significantly higher than the background fluctuations. In this section we study the probability of GRB detection by arrays of WCD placed at Chacaltaya, Pico Espejo and Sierra Negra individually. The flux of a GRB at Earth can be described by the power law (See 2006 BATSE catalog \cite{Kaneko:2006qe})
{\small
\begin{equation}
J(E)=K \left(\frac{E}{E_0}\right)^{-\alpha},
\end{equation}
}   
\noindent where $K$ is a spectrum normalization constant, $E_0$ is fixed to $1$GeV and the spectral index $\alpha$ takes values  between $2$ and $3$. The mean value of $\alpha$ is around $2.2$ and no high energy cut-off has been observed \cite{JPreece122, RBriggs524}. We will naively assume that such power law is still valid for high energy photons. We compute the mean number of triggered particles generated during a GRB integrating the typical GRB flux and the mean number of triggered particles from a shower generated by a photon with energy ($<N(E)_{trig}>$)over the primary energies. 
{\small
\begin{equation}\label{Ntriggalpha}
<n_{trigger}(\alpha)>=\int_{0.5 \mbox{GeV}}^{1\mbox{TeV}}J(E)<n_{trig}(E)>dE\; .
\end{equation}
}   
In this calculation, $K$ is still a free parameter. Minimal $K$ can be calculated from the condition:
{\small
\begin{equation}\label{background}
<n_{trigger}(\alpha)>\; \geq\; \frac{s}{\cos{(\theta) A}}\sqrt{\frac{B}{\Delta t N_d}}, 
\end{equation}}
\noindent here $s$ is the statistical significance, $\theta$ is the zenith angle, $B$ is the background counting rate, $\Delta t$ is the duration of the flux of photons, $A$ is the area of the WCD and $N_d$ is the number of WCDs at each site. We will take $\Delta t=1 s$, $s=5$, as we are working with vertical showers for the tanks response then $\theta=0$, $B=12$ kHz per tank (which is the background at Chacaltaya), our tanks have typically $A=4$ m$^2$ and $N_d=3$. Once minimal $K(\alpha)$ is calculated, sweeping on various values of the spectral parameter, then one can estimate the properties of a possible GRB which can be detected by our surface detectors. The GRB flux at Earth can be calculated for different values of $\alpha$ from its luminosity $L$ at certain redshift $z$ in a range of energies and the distance from the source $r(z)$ as follows
{\small
\begin{equation}\label{GRBflux}
\phi(\alpha)=\frac{L}{4\pi r^2(z)}=\int_{\frac{E_{min}}{1+z}}^{\frac{E_{max}}{1+z}} J(E) E dE\; .
\end{equation}
} 
An estimation about the minimal fluences that we could expect to detect with our surface detectors can be calculated by the integral in \ref{GRBflux} from $0.5$ GeV to $E_{max}$, where $E_{max}$ varies from 1 GeV to 1 TeV. Results of these GRB fluences as function of the spectral parameter $\alpha$ are represented in figure \ref{fig:Fluencevsalpha} for $E_{max}=1,\;10\;,100\; \mbox{and}\; 1000$ GeV. We used $\Delta t=1$ s. We take $z\ll 1$ supposing small distortions in the spectrum. Fluences shown in figure \ref{fig:Fluencevsalpha} exhibit that our surface detectors seems to be sensitive in the range of $10^{-7}$ and $10^{-6}$ erg$/$cm$^2$.\\ 
These results are comparable with the prediction made in \cite{Vernetto:1999jz} for high altitude ($4.5$ Km a.s.l. to $5.5$ Km a.s.l.) surface detectors of $10^3$ m$^2$ scintillators array. It is reasonable to expect sensitivity to even smaller fluences since we use WCDs  able to detect the huge amount of secondary photons coming from high energy primary photons. The Pierre Auger Observatory shows sensitivity to high energy (between $10$ MeV  to $1$ TeV) GRB photon fluences, using the single particle technique, between $10^{-6}$ and $10^{-3}$ erg$/$cm$^2$ \cite{Allard}. BATSE detectors were reported\cite{Kaneko:2006qe} to be sensitive to fluences of $\sim 10^{-8}$ in the rage of energies from $25$KeV to $2$MeV among $2704$ events identified as GRBs.       

\section{Discussion and conclusions} 
Our atmospheric cascade simulations showed significant differences for zenith angles between $0^0$ and $30^0$. Since high altitude surface detectors would be very sensitive to injection angle, inclusion of slanted showers in the response of the WCD and the extension of this study to larger zenith angles ($>30^\circ$) would be relevant. On the other hand, shower simulation results were not sensible to the geomagnetic field for the sites considered (with different latitudes). Nevertheless, this has to be checked for greater zenith angles \cite{EnPreparacion}.\\
Detection of GRBs with WCDs is powered by the presence of high energy pairs $e^+/e^-$ into the water, created by photons ($E_\gamma > 1.55\, \rm MeV$), able to produce Cherenkov light. Our results showed a significant amount of particles arriving with energies close to $10$ MeV even for primary energies of $1$ GeV. Since the number of photons arriving to the detectors is about one order of magnitude higher than charged particles, their contribution to the signal is valuable.
Moreover, the three sites studied in this work, Chacaltaya, Sierra Negra and Pico Espejo have a notable potential on GRB high energy photons detection, using the single particle technique, due its high altitude. In the group, Chacaltaya (5200 m a.s.l.,  $\sim 640$ m higher than Sierra Negra and $\sim 440$ m higher than Pico Espejo) highlights with a minimal fluence lower than Pico Espejo and Sierra Negra by an order of magnitude, meanwhile detectable fluences appear very close for the last two (for example, Pico Espejo is $0.1$ lower than Sierra Negra for $E_{max}=1\, \rm GeV$ and $\alpha=2.3$). Our analisys showed that the LAGO high altitude sites (with $3$ tanks at each site) are as or more sensitive for GRB detection than the complete Pierre Auger Observatory ($1600$ tanks)\\
This results will allow us to analyze the space of parameters involved in GRB detection, using the single particle technique, and propose limits in order to compare with satellite observations \cite{1413}

\section*{Acknowledgments}
LAGO collaboration is very thankful to the Pierre Auger collaboration for providing us the PMTs and electronic equipment. A. D. Castro would like to thank {\it Laboratorio de Computaci\'on de Alto Rendimiento-USB} specially to Yudith Cardinale, Leonardo Quevedo and Jes\'us De Oliveira for computational resources and technical support. This work was partially supported by the Grant 5681-08-FUNINDES-USB-LOCTI. 





\begin{thebibliography}{99}
  
  

 \bibitem{Meszaros:2006rc}
  P.~Meszaros,
  Rept.\ Prog.\ Phys.\  {\bf 69} (2006) 2259
  [arXiv:astro-ph/0605208].
  
  \bibitem{Kaneko:2006qe}
  Y.~Kaneko, R.~D.~Preece, M.~S.~Briggs, W.~S.~Paciesas, C.~A.~Meegan and D.~L.~Band,
  arXiv:astro-ph/0605427.
  
  
\bibitem{Meszaros:1993ft}
  P.~Meszaros, M.~J.~Rees and H.~Papathanassiou,
  Astrophys.\ J.\  {\bf 432} (1994) 181
  [arXiv:astro-ph/9311071].
  
  
\bibitem{Vietri:1997st}
  M.~Vietri,
  Phys.\ Rev.\ Lett.\  {\bf 78} (1997) 4328
  [arXiv:astro-ph/9705061].
  
  
\bibitem{Baring:1997tv}
  M.~G.~Baring,
  arXiv:astro-ph/9711256.
  
  \bibitem{Castellina:2000jq}
  A.~Castellina {\it et al.}  [INCA Collaboration],
  Nuovo Cim.\  {\bf 24C} (2001) 631
  [arXiv:astro-ph/0011241].


\bibitem{Vernetto:1999jz}
  S.~Vernetto,
  Astropart.\ Phys.\  {\bf 13} (2000) 75
  [arXiv:astro-ph/9904324].

\bibitem{Allard:2008zz}
  D.~Allard {\it et al.},
  Nucl.\ Instrum.\ Meth.\  A {\bf 595} (2008) 70.


\bibitem{Bertou:2007hh}
  X.~Bertou  [Pierre Auger Collaboration],
  arXiv:0706.1256 [astro-ph].

\bibitem{1413}
  X.~Bertou~for~the~LAGO~Collaboration,
  ICRC{\bf1413} (2009).


\bibitem{aires}
   S. J. Sciutto,
   ``AIRES, A system for air shower simulations,''
   VERSION {\bf 2.6.0} (2002) 
  

\bibitem{EnPreparacion}
  LAGO~Collaboration,
  In Preparation.

\bibitem{1417}
  H.~Salazar~for~the~LAGO~Collaboration,
  ICRC{\bf 1417} (2009).




\bibitem{JPreece122}
  J.~Preece et. al,
  Ap\ J. {\bf 122} (1999) 465.

\bibitem{RBriggs524}
  R.~Briggs et. al,
  Ap\ J. {\bf 524} (1999) 82.

 \bibitem{Allard}
  D.~Allard {\it et al.}  [LAGO Collaboration],
  ICRC at Pune 2005.\  {\bf 00} (2005) 101-104
 





 
 
 
\end{thebibliography}
\end{document}